\documentclass[12pt,final]{iopart}
\usepackage{epsfig}
\usepackage{citesort}

\usepackage{a4wide}

\newcommand{\gtrsim}{\,\rlap{\lower3.7pt\hbox{$\mathchar\sim$}}
\raise1pt\hbox{$>$}\,}
\newcommand{\lesssim}{\,\rlap{\lower3.7pt\hbox{$\mathchar\sim$}}
\raise1pt\hbox{$<$}\,}

\usepackage{graphicx}
\usepackage[]{subfigure}
\usepackage{epsfig}

\newcommand{\nc}{\newcommand}

\nc{\be}[1]{\begin{equation}\mbox{$\label{#1}$}}
\nc{\bea}[1]{\begin{eqnarray} \mbox{$\label{#1}$}}
\nc{\Section}[2]{\section{#2}\label{#1}}
\nc{\Bibitem}[1]{\bibitem{#1}}
\nc{\Label}[1]{\label{#1}}

\nc{\eea}{\end{eqnarray}}
\nc{\ee}{\end{equation}}

\nc{\bdm}{\begin{displaymath}}
\nc{\edm}{\end{displaymath}}
\nc{\dpsty}{\displaystyle}
\nc{\bc}{\begin{center}}
\nc{\ec}{\end{center}}
\nc{\ba}{\begin{array}}
\nc{\ea}{\end{array}}
\nc{\bab}{\begin{abstract}}
\nc{\eab}{\end{abstract}}
\nc{\btab}{\begin{tabular}}
\nc{\etab}{\end{tabular}}
\nc{\bit}{\begin{itemize}}
\nc{\eit}{\end{itemize}}
\nc{\ben}{\begin{enumerate}}
\nc{\een}{\end{enumerate}}
\nc{\bfig}{\begin{figure}}
\nc{\efig}{\end{figure}}

\nc{\arreq}{&\!=\!&}
\nc{\arrmi}{&\!-\!&}
\nc{\arrpl}{&\!+\!&}
\nc{\arrap}{&\!\!\!\approx\!\!\!&}
\nc{\non}{\nonumber}
\nc{\align}{\!\!\!\!\!\!\!\!&&}

\def\lsim{\; \raise0.3ex\hbox{$<$\kern-0.75em
      \raise-1.1ex\hbox{$\sim$}}\; }
\def\gsim{\; \raise0.3ex\hbox{$>$\kern-0.75em
      \raise-1.1ex\hbox{$\sim$}}\; }

\nc{\DOT}{\hspace{-0.08in}{\bf .}\hspace{0.1in}}
\nc{\Laada}{\hbox {$\sqcap$ \kern -1em $\sqcup$}}
\nc\loota{{\scriptstyle\sqcap\kern-0.55em\hbox{$\scriptstyle\sqcup$}}}
\nc\Loota{{\sqcap\kern-0.65em\hbox{$\sqcup$}}}
\nc\laada{\Loota}
\nc{\qed}{\hskip 3em \hbox{\BOX} \vskip 2ex}

\nc{\real}{{\rm I \! R}}
\nc{\Z}{{\sf Z \!\!\! Z}}
\nc{\complex}{{\rm C\!\!\! {\sf I}\,\,}}
\def\bigid{\leavevmode\hbox{\small1\kern-3.8pt\normalsize1}}
\def\id{\leavevmode\hbox{\small1\kern-3.3pt\normalsize1}}
\nc{\slask}{\!\!\!/}
\nc{\bis}{{\prime\prime}}
\nc{\pa}{\partial}
\nc{\na}{\nabla}
\nc{\ra}{\rangle}
\nc{\la}{\langle}
\nc{\goto}{\rightarrow}
\nc{\swap}{\leftrightarrow}

\nc{\EE}[1]{ \mbox{$\cdot10^{#1}$} }
\nc{\abs}[1]{\left|#1\right|}
\nc{\at}[2]{\left.#1\right|_{#2}}
\nc{\norm}[1]{\|#1\|}
\nc{\abscut}[2]{\Abs{#1}_{\scriptscriptstyle#2}}
\nc{\vek}[1]{{\rm\bf #1}}
\nc{\integral}[2]{\int\limits_{#1}^{#2}}
\nc{\inv}[1]{\frac{1}{#1}}
\nc{\dd}[2]{{{\partial #1}\over{\partial #2}}}
\nc{\ddd}[2]{{{{\partial}^2 #1}\over{\partial {#2}^2}}}
\nc{\dddd}[3]{{{{\partial}^2 #1}\over
    {\partial #2 \partial #3}}}
\nc{\dder}[2]{{{d #1}\over{d #2}}}
\nc{\ddder}[2]{{{d^2 #1}\over{d {#2}^2}}}
\nc{\dddder}[3]{{d^2 #1}\over
    {d #2 d #3}}
\nc{\dx}[1]{d\,^{#1}x}
\nc{\dy}[1]{d\,^{#1}y}
\nc{\dz}[1]{d\,^{#1}z}
\nc{\dl}[1]{\frac{d\,^{#1}l}{(2\pi)^{#1}}}
\nc{\dk}[1]{\frac{d\,^{#1}k}{(2\pi)^{#1}}}
\nc{\dq}[1]{\frac{d\,^{#1}q}{(2\pi)^{#1}}}

\nc{\bfT}{{\bf T }}

\nc{\cA}{{\cal A}}
\nc{\cB}{{\cal B}}
\nc{\cD}{{\cal D}}
\nc{\cE}{{\cal E}}
\nc{\cG}{{\cal G}}
\nc{\cH}{{\cal H}}
\nc{\cL}{{\cal L}}
\nc{\cO}{{\cal O}}
\nc{\cT}{{\cal T}}
\nc{\cN}{{\cal N}}
\nc{\cR}{{\cal R}}

\nc{\rvac}[1]{|{\cal O}#1\rangle}
\nc{\lvac}[1]{\langle{\cal O}#1|}
\nc{\rvacb}[1]{|{\cal O}_\beta #1\rangle}
\nc{\lvacb}[1]{\langle{\cal O}_\beta #1 |}
\nc{\bb}{\bar{\beta}}
\nc{\bt}{\tilde{\beta}}
\nc{\ctH}{\tilde{\cal H}}
\nc{\chH}{\hat{\cal H}}
\nc{\al}{\alpha}
\nc{\g}{\gamma}
\nc{\Del}{\Delta}
\nc{\eps}{\epsilon}
\nc{\lam}{\lambda}
\nc{\Om}{\Omega}
\nc{\ve}{\varepsilon}
\nc{\mn}{{\mu\nu}}
\nc{\vp}{\varphi}

\nc{\rf}[1]{(\ref{#1})}
\nc{\nn}{\nonumber \\*}
\nc{\bfB}{\bf{B}}
\nc{\bfv}{\bf{v}}
\nc{\bfx}{\bf{x}}
\nc{\bfy}{\bf{y}}
\nc{\vx}{\vec{x}}
\nc{\vy}{\vec{y}}
\nc{\oB}{\overline{B}}
\nc{\oI}{\overline{I}}
\nc{\oR}{\overline{R}}
\nc{\rar}{\rightarrow}
\nc{\ti}{\times}
\nc{\slsh}{\hskip-5pt/}
\nc{\sm}{Standard~Model~}
\nc{\MP}{M_{\rm Pl}}
\nc{\mpl}{M_{\rm Pl}}
\nc{\tp}{t_{\rm Pl}}

\nc{\pmin}{p_{\rm min}}
\nc{\pmax}{p_{\rm max}}
\nc{\fo}{f_0}
\nc{\foi}{f_{0,i}\,}
\nc{\fop}{f_0^P}
\nc{\fou}{f_0^U}

\nc{\eff}{{\rm eff}}
\nc{\MT}{M_{\rm T}}
\nc{\ML}{M_{\rm L}}
\nc{\kk}{\vek{k}}
\nc{\pp}{{\rm p}}
\nc{\half}{{1\over 2}}
\nc{\w}{\omega}
\nc{\uhat}{\hat{U}_\w}

\nc{\ie}{{\it i.e. }}
\nc{\eg}{{\it e.g. }}
\nc{\trh}{T_{\rm RH}}
\nc{\ad}{{a'\over a}}
\nc{\bd}{{b'\over b}}
\nc{\Rd}{{R'\over R}}
\nc{\diag}{{\textrm{diag}}}
\nc{\mato}[1]{\tilde{#1}}
\nc{\sinn}{\textrm{sinn}}
\nc{\sech}{\textrm{sech}}
\nc{\I}{\textrm{I}}
\nc{\II}{\textrm{II}}
\nc{\III}{\textrm{III}}
\nc{\vev}[1]{\langle #1 \rangle}
\nc{\hyp}{\,\; F_{1{\hskip -16pt}2}{\hskip 11pt}}
\nc{\brhom}{\overline{\rho}_M}
\nc{\brho}{\overline{\rho}}
\nc{\rhob}{\overline{\rho}}
\nc{\Pb}{\overline{P}}
\nc{\bH}{\overline{H}}
\nc{\ep}{{1+4\eps}}

\nc{\deriv}[2]{ 
\frac{\mathrm{d}#1}{\mathrm{d}#2}
}
\nc{\Mnu}{M_\nu}
\nc{\bee}{\begin{equation}}
\nc{\ene}{\end{equation}}
\nc{\hdp}{\sigma_8 (\Omega_{\rm m}/0.3)^{0.37}}
\nc{\avis}{\alpha_{vis}}
\nc{\cvis}{c^2_{vis}}
\nc{\clam}{c^2_{lam}}
\nc{\mnue}{m_{\nu_e}}

\def\smiley{\hbox{\large$\bigcirc$\hspace{-.80em}%
\raise.2ex\hbox{$\cdot\cdot$}\kern-.61em    %--- .56
\lower.2ex\hbox{\scriptsize$\smile$}}\ }

\def\frowney{\hbox{\large$\bigcirc$\hspace{-.80em}%
\raise.2ex\hbox{$\cdot\cdot$}\kern-.635em
\lower.2ex\hbox{\scriptsize$\frown$}}\ }

\begin{document}

\title{Cosmological implications of the KATRIN experiment}
\author{Jostein R. Kristiansen}
\mailto{j.r.kristiansen@astro.uio.no}
\address{Institute of Theoretical Astrophysics, University of Oslo, Box 1029, 0315 Oslo, Norway}
\author{\O{}ystein Elgar\o{}y}
\mailto{oelgaroy@astro.uio.no}
\address{Institute of Theoretical Astrophysics, University of Oslo, Box 1029, 0315 Oslo, Norway}

\date{\today}

\begin{abstract}
The upcoming Karlsruhe Tritium Neutrino (KATRIN) experiment will put
unprecedented constraints on the absolute mass of the electron
neutrino, $\mnue$. In this paper we investigate how this information on $\mnue$ will affect our constraints on cosmological
parameters. We consider two scenarios; one where $\mnue=0$ (i.e., no detection by KATRIN), 
and one
where $\mnue=0.3$eV. We find that the constraints on $\mnue$ from
KATRIN will affect estimates of some important  cosmological
parameters significantly. For example, the significance of
$n_s<1$ and the inferred value of $\Omega_\Lambda$ depend on the
results from the KATRIN experiment. 

\end{abstract}

%\pacs{98.80.-k,98.80.Jk}

\maketitle

\section{Introduction}   
The large amount of new cosmological data in the last decade has lead
to what one may call the cosmological standard model. In this model the universe is
close to flat, homogeneous and isotropic on sufficiently large scales, and today the
energy density of the universe is dominated by dark energy ($\sim
74\%$), dark matter ($\sim 22\%$) and baryonic matter ($\sim 4\%$).
This model is consistent with data ranging from the WMAP
measurements of the anisotropies of the cosmic microwave background
(CMB) radiation \cite{spergel:2006} to observations of supernovae of  type
1a, galaxy distributions and several other observables (with a few
exceptions, see \cite{lieu:2007}). 
It is often claimed that most of the data can be fitted with only
six free parameters. This claim rests on the assumption of 
massless neutrinos, an assumption justified by the fact that adding the sum 
of the neutrino masses as a
free parameter does not improve the fit substantially. 

However, from
the observation of neutrino oscillations, there is a compelling body
of evidence for non-zero neutrino masses (see \cite{messier:2006} for
a review). Oscillation experiments do not give us any information on
the absolute mass scale of neutrinos, only on the mass differences
between the different mass eigenstates and mixing angles. The current best
upper bounds on the neutrino mass from particle experiments come from
the Troitsk \cite{lobashev:1999} and Mainz \cite{weinheimer:1999}
tritium beta decay experiments that found upper bounds on $\mnue$ of
$\mnue<2.2$eV (95\% C.L.). The KATRIN
experiment \cite{katrin:2004} that will start taking data in 2010, is expected to lower
this limit on $\mnue$ by an order of magnitude to $\mnue<0.2$eV (in
the case of no detection) after three years of running.   

Effects of neutrino masses can also be seen in cosmological
observables, and the best upper limits on the absolute scale of the neutrino mass today
come from cosmology. Both CMB and the large scale structures (LSS) of
the galaxy distribution are probes that are sensitive to the neutrino
mass, the observable quantity being the sum of the three neutrino mass
eigenstates, $\Mnu = \sum_i m_{\nu_i}$. The upper bounds on $\Mnu$ from cosmology range
from $\Mnu<0.2$eV \cite{seljak:2006} to $\Mnu<2.0$eV
\cite{spergel:2006} (95\% C.L.), depending on the data
\cite{kristiansen:2007} and cosmological model \cite{zunckel:2006}
used. 

On the experimental side there is a claim of detection of the
absolute scale of the neutrino mass from the Heidelberg-Moscow
neutrinoless double beta decay experiment, with an effective electron
neutrino mass of $\langle \mnue \rangle = (0.2 - 0.6)$eV (99.73\% C.L.)
\cite{klapdor:2004}. However, these results are regarded somewhat
controversial. The cosmological implications of this result are
discussed in \cite{macorra:2006}. 

We know that neutrinos are massive, and since we have
no current priors on the neutrino mass in the allowed cosmological
range, one should always marginalize over $\Mnu$ when constraining
other cosmological parameters. $\Mnu$ turns out to be partially
degenerate with several of the standard cosmological parameters,
such that this marginalization over $\Mnu$ weakens the bounds on the
other parameters in our model. Thus, any prior knowledge of $\Mnu$
from non-cosmological experiments will serve to break degeneracies and
improve the constraints on other cosmological parameters. The KATRIN
experiment will provide us with such a prior on $\Mnu$ in a range that
is relevant for cosmology. In this paper we investigate how the
results from KATRIN will affect our estimates of other cosmological
parameters. 

Limits on the neutrino mass  when combining results from KATRIN
and WMAP
have been studied in a recent paper by  that H{\o}st et al. \cite{host:2007}. Our emphasis in this paper is on how other cosmological  
parameters are affected when the results from KATRIN are used as an 
external prior.  

Section \ref{sec:cosmology} contains a short review on the effect of massive neutrinos
in cosmology. In section \ref{sec:data}
we will present the data and methods that we will use in our analysis,
including the assumed priors from the KATRIN experiment. Then we will
present our results in section \ref{sec:results}. A comparison of
$\chi^2$ values found when introducing the KATRIN priors is presented
in section \ref{sec:chisq}. Finally  we summarize and conclude
in section \ref{sec:conclusions}. 

\section{Cosmology with massive neutrinos}
\label{sec:cosmology}
All our results are derived within the standard cosmological
paradigm of a flat $\Lambda$CDM model, using the following free
parameters: $\{ \Omega_bh^2, \Omega_m, \log(10^{10}A_S), h, n_s, \tau,
M_\nu\}$. Here $\Omega_i$ denotes the energy density of energy
component $i$ (m=matter, b=baryons, $\Lambda$=cosmological constant,
CDM=cold dark matter) relative to the total energy density of a spatially flat 
universe. The matter density, $\Omega_m$, is the sum of all non-relativistic components,
such that $\Omega_m=\Omega_{CDM}+\Omega_b+\Omega_\nu$. The parameter $h$ is the 
dimensionless Hubble parameter, defined by $H_0 =
100h \textrm{km s}^{-1}\textrm{Mpc}^{-1}$, $A_s$ denotes the amplitude
of the primordial fluctuations, while $n_s$ gives the tilt of the
primordial power spectrum. Finally, $\tau$ is the optical depth at
reionization. For more details on the parameter definitions, see the description of the 
CosmoMC code \cite{lewis:2002}. The effect of massive neutrinos on cosmological observables 
is parameterized by $M_\nu$, the sum of the neutrino masses, and is
related to the neutrino energy density by the simple relation
\cite{lesgourgues:2006} $\Omega_\nu h^2 =
\frac{\Mnu}{93.14\textrm{eV}}$. We will also extend the parameter space by including $w$, the
equation of state parameter of dark energy, as a free parameter. We
will assume $w$ to be
constant. This parameter may be interesting to study, as it is
fundamental in the understanding of the nature of dark energy, and
since it is known to be slightly correlated with $\Mnu$
\cite{hannestad:2005}. It should be stressed that this
analysis rests on the assumption of a standard thermal background of 3 weakly
interacting neutrino species. Alternatives to this picture are studied in
e.g. \cite{beacom:2004} and \cite{bell:2005}.

Recent reviews of the role of massive neutrinos in cosmology can be found
in \cite{elgaroy:2005,lesgourgues:2006}. In this section we will only give a brief
description of the most important effects of $\Mnu$ on relevant
cosmological observables. We will throughout this work assume that the
neutrino mass eigenstates are degenerate, such that
$\Mnu=3\mnue$. In the mass range that we will operating in, it has
been shown that this is a valid simplification when it comes to
cosmological observables \cite{slosar:2006}.

\begin{figure}
\center
\includegraphics[width=8cm]{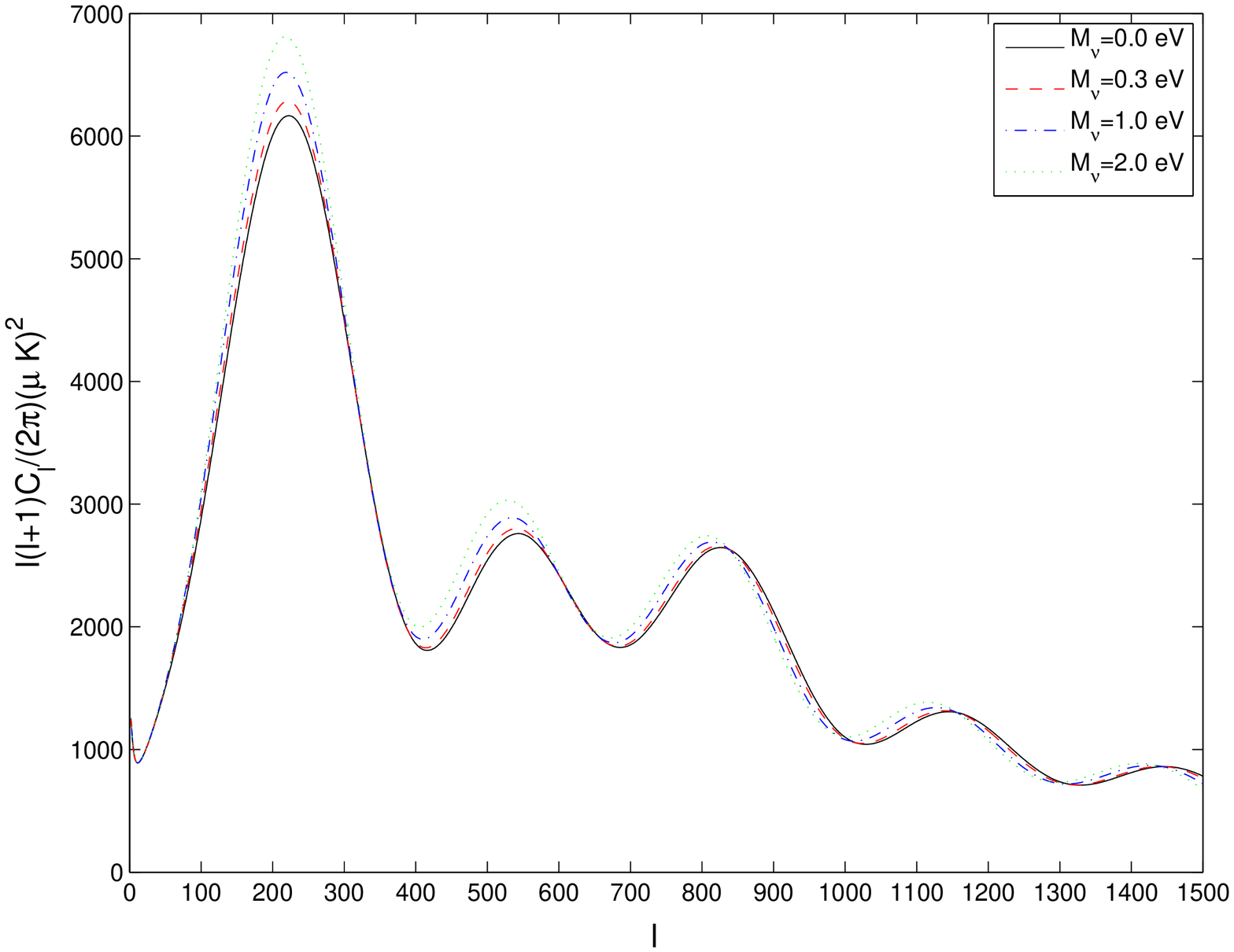}
\includegraphics[width=8cm]{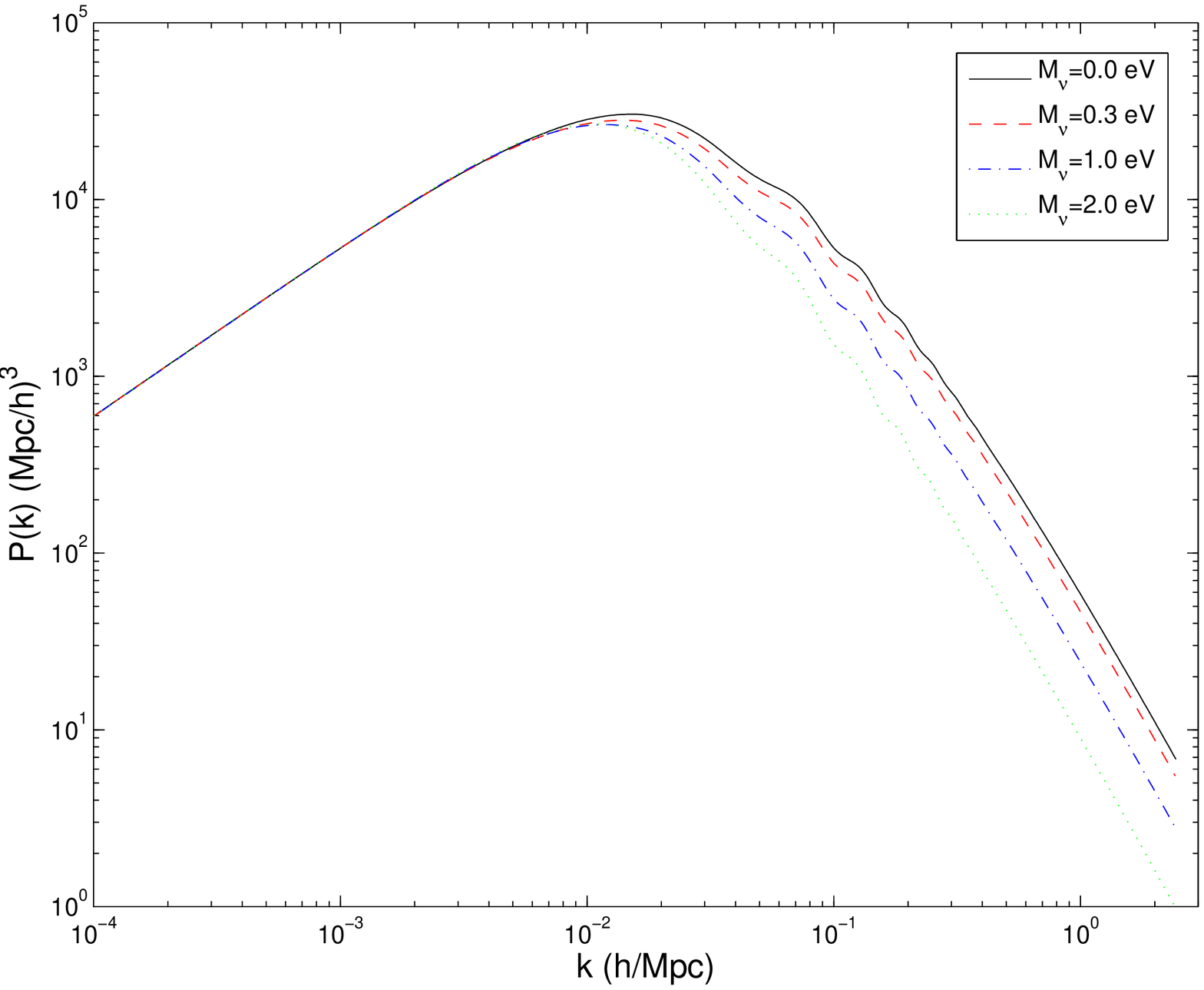}
\caption{\label{fig:P} CMB (left panel) and LSS (right panel) power
  spectra with different values of $\Mnu$. Here $\Omega_m$ is held
  constant, and increasing $\Mnu$ has been compensated by decreasing
  $\Omega_{CDM}$ correspondingly.}
\end{figure}

Effects on the CMB from massive neutrinos manifest themselves mainly on the level of background
evolution. In the neutrino mass ranges relevant to us, the neutrinos
will still be relativistic at the time of matter-radiation
equality, and must be regarded as a radiation component when it comes
to the background evolution of the universe. Increasing $\Mnu$ (and
thus also $\Omega_\nu$), keeping $\Omega_m$
constant, will thus postpone the time of matter-radiation equality.
This will enhance the acoustic peaks in the CMB power spectrum and
give a small horizontal shift of the peaks to larger scales. This
effect is shown in the left panel of Figure \ref{fig:P}. To compensate
for this effect, one can increase $\Omega_m$ and decrease $H_0$. It is
already obvious that $\Mnu$ will be correlated with both $H_0$ and $\Omega_m$ (and
thus also $\Omega_\Lambda$ when we stick to the assumption of spatial
flatness). Another effect comes from neutrino free-streaming, which
will smoothen out gravitational wells on scales below an $M_\nu$-
dependent neutrino free streaming scale \cite{dodelson:1996,ichikawa:2005}. On scales smaller than this, the acoustic
oscillations will be enhanced, increasing the height of the peaks in
the CMB power spectrum. 

Neutrino masses affect the LSS power spectrum in an even more distinct
way. Again, massive neutrinos will suppress structure
growth on scales below a free streaming scale given by
\cite{lesgourgues:2006}
\begin{equation}
k_{nr} = 0.010 \sqrt{\frac{M_\nu \Omega_m}{1 \textrm{eV}}}h \textrm{Mpc}^{-1}.
\end{equation}
The smaller $\Mnu$, the larger scales will be affected, and the larger
$\Mnu$, the more suppression of power on the scales affected. The
effect of neutrino mass on the matter power spectrum can be seen in
Figure \ref{fig:P}. Again, $\Omega_m$ is kept constant, and increasing
$\Mnu$ has been compensated for by decreasing $\Omega_{CDM}$ correspondingly.

\section{Data and methods}
\label{sec:data}

\subsection{Cosmological data}
Our analysis include both observations of CMB, LSS, SN1a,
information about baryonic acoustic oscillations (BAO) in the matter power
spectrum and constraints from the cluster mass function from weak
gravitational lensing. We have also applied priors on $H_0$ and
$\Omega_b$.

The CMB data used in our analysis, comes from the temperature \cite{hinshaw:2006}  and
polarization \cite{page:2006} data
from the 3-year data release from the WMAP team. The WMAP experiment is
a satellite based full-sky survey of the CMB temperature anisotropies
and polarization. In our analysis of the WMAP data we have used the
Fortran 90 likelihood code\footnote{http://lambda.gsfc.nasa.gov;
  version v2p2} provided with the data release.

We have used LSS data from the Sloan Digital Sky Survey (SDSS)
luminous red galaxy (LRG) sample \cite{tegmark:2006}. As SN1a data we
have used the sample from the Supernova Legacy Survey (SNLS)
\cite{astier:2006}. Other probes of the matter distribution that we have applied come from
the measurement of the baryonic acoustic peak (BAO) in the matter
power spectrum and the cluster mass function (CMF) from weak gravitational
lensing. The BAO constraint comes from the SDSS-LRG sample
\cite{eisenstein:2005}, and we have adopted the fit function from
\cite{goobar:2006}, 
\bee
A_{\textrm{BAO}} = 0.469 \left( \frac{n_s}{0.98} \right)^{-0.35} (1+0.94 f_\nu) \pm 0.017,
\ene
where
\bee
A_{\textrm{BAO}} \equiv \left[ D_M(z)^2 \frac{z}{H(z)} \right]^{1/3}
\frac{\sqrt{\Omega_m H_0^2}}{z},
\ene
and $D_M(z)$ is the comoving angular diameter distance at redshift
$z$. 

Handles on parameters governing the clustering of matter are also
provided by the cluster mass function. The cluster mass function from weak gravitational lensing, as measured
in \cite{dahle:2006}, gives constraints on a combination of $\Omega_m$
and $\sigma_8$ (the root-mean-square  mass fluctuations in spheres of
radius $8h^{-1}$Mpc). We have adopted the fit-function for $\chi^2_{\textrm{CMF}}$ from
\cite{kristiansen:2007},
\bee
\chi^2_\textrm{CMF}=10000u^4 + 6726u^3+1230u^2-4.09u+0.004
\ene
where $u=\sigma_8(\Omega_m/0.3)^{0.37}-0.67$. 

A prior on the Hubble parameter, $h=0.72 \pm 0.08$
\cite{freedman:2001} comes from the Hubble Space Telescope (HST) key
project. From big bang nucleosyntesis (BBN) we adopt a prior on the
baryon density today, $\Omega_b h^2 = 0.022 \pm 0.002$
\cite{burles:2001,cyburt:2004,serpico:2004}. 

Throughout the entire work we also apply a top-hat prior on the age $t_0$ of
the universe:  $10\;{\rm Gyr}<t_0<20\;{\rm Gyr}$.

\subsection{Constraints from KATRIN}
\label{sec:KATRIN}
The KATRIN \cite{katrin:2004} experiment measures the energy
distribution of electrons from tritium beta decay. The exact shape of
the end of this spectrum will depend on how much of the energy that is
bound in the outgoing electron neutrinos, and thus also be a probe of
the electron neutrino mass. If KATRIN does not detect $\mnue$, they
are expected to place an upper limit on $\mnue<0.2$eV (90\%
C.L.). They expect to reach an uncertainty of $\sigma_{\mnue^2}
\approx 0.025\textrm{eV}^2$. 

Here we have adopted this
uncertainty for two cases, one assuming $\mnue=0$eV (i.e., no detection by KATRIN), 
and one assuming
$\mnue=0.3$eV (giving $\Mnu=0.9$eV). Further we have assumed the
Gaussian distribution of $\mnue^2$ around these values
\cite{katrin:2004,host:2007}, and used this
as a prior in our cosmological parameter analysis.

\subsection{Method}
Employing the publicly available Markov Chain
Monte Carlo code CosmoMC\cite{lewis:2002} we have studied our seven-parameter 
model for two combinations of datasets; first using only
WMAP data, and then adding LSS, SN1a data and priors from HST, BBN, BAO and
CMF. In both cases we have compared the results from using only
cosmological data, and from adding priors from KATRIN in the case of
$\mnue=0$eV and $\mnue=0.3$eV. First we will assume $w=-1$
(cosmological constant). 

Yet more freedom in the cosmological model might be added by including
$w$ as a free parameter, yielding a more general form of the
dark energy component. We will also include $w$ in our analysis, assuming it to
be constant. 

\section{Results}
\label{sec:results}

\subsection{A 7-parameter model}
\begin{figure}[htb]
\center
\includegraphics[width=14cm]{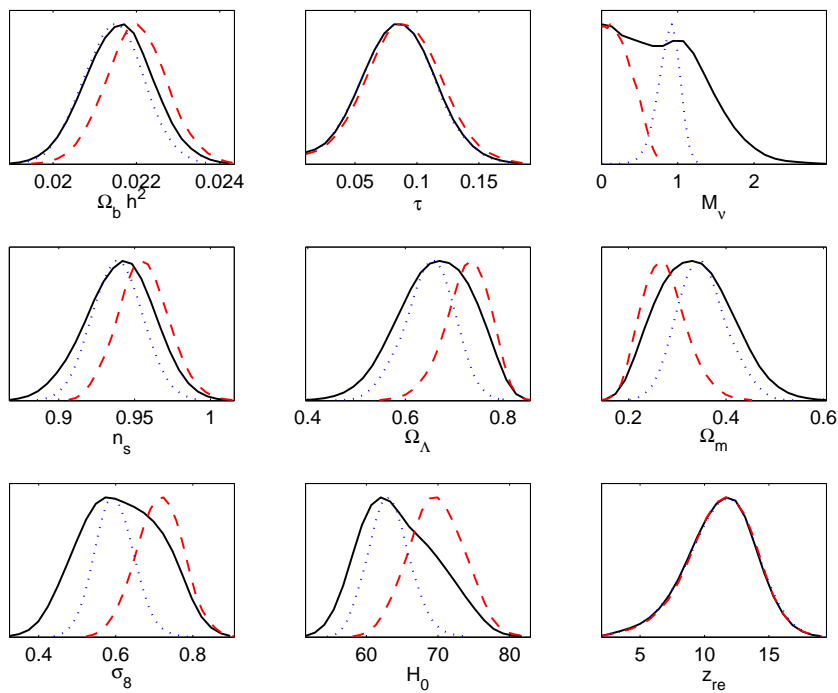}
\caption{\label{fig:1} Marginalized parameter distributions when using
WMAP data (black solid lines), compared to the resulting distributions when
adding KATRIN data with $\mnue=0$eV (red dashed lines) and $\mnue=0.3$eV
(blue dotted lines).}
\end{figure}

Starting out, we considered the simplest case using only the standard 7
parameter universe model, as explained in section
\ref{sec:cosmology}, and WMAP data only. Then we added the assumed
KATRIN priors for $\mnue=0$eV and $\mnue=0.3$eV as explained in
section \ref{sec:KATRIN}. The results are summarized in Figure
\ref{fig:1} and Table \ref{tab:1}. 
\begin{table}[htb]
\center
\begin{tabular}{cccc}
\hline \hline
%Parameter & \multicolumn{3}{c}{WMAP} & \multicolumn{3}{c}{WMAP+LSS+SN1a+HST+BBN+CMF} \\
%\hline \hline
&  $\mnue = \textrm{free}$ & $\mnue=0$eV & $\mnue=0.3$eV \\
\hline
$\Omega_b h^2$ & $0.0216\pm0.0008$ & $0.0220\pm0.0007$ & $0.0215\pm0.0007$ \\
$h$ & $0.65^{+0.06}_{-0.05}$ & $0.70\pm+0.04$ & $0.63\pm0.03$\\
$\tau$ & $0.085\pm0.029$ & $0.089\pm0.029$ & $0.085\pm0.028$\\
$n_s$ & $0.940\pm0.022$ & $0.956\pm0.016$ & $0.939\pm0.016$\\
$\sigma_8$ & $0.61^{+0.09}_{-0.10}$ & $0.71\pm0.06$ & $0.60\pm0.05$ \\
$\Omega_\Lambda$ & $0.66\pm0.07$ & $0.73\pm0.04$ & $0.65\pm0.05$\\
$\Omega_m$ & $0.34\pm0.07$& $0.27\pm0.04$ & $0.35\pm0.05$ \\
$z_\textrm{re}$ & $11.2\pm2.6$ & $11.3\pm2.6$ & $11.2\pm2.7$\\
$\Mnu$ (eV) & $<1.75$ (95\%C.L.)& $<0.58$ (95\% C.L.)& $0.87^{+0.10}_{-0.13}$\\
\hline
\end{tabular}
\caption{\label{tab:1}Limits on cosmological parameters with different
  priors on the neutrino mass when using WMAP data only. In the left column are the results from
having no priors on $\Mnu$ (black solid lines in Figure \ref{fig:1}),
the middle column shows the results when using the assumed KATRIN prior in
the case of $\mnue=0$eV (red dashed lines in Figure \ref{fig:1}), and
the rightmost column gives the results with an assumed KATRIN prior
for $\mnue=0.3$eV.}
\end{table}
One easily sees that the different priors from KATRIN indeed affect
some of the parameter constraints. 

Evidently the inferred value of $h$ depend
heavily on which prior we assume on $\mnue$. A larger $\Mnu$ requires
a smaller $h$. This degeneracy is not very surprising, as both $\Mnu$
and $H_0$ tend to shift the positions of the acoustic peaks in the CMB
power spectrum, as mentioned in section \ref{sec:cosmology}. Note that
we have assumed no prior on $H_0$ in this case. Also $\sigma_8$ is highly dependent on the priors on $\mnue$. We saw
in section \ref{sec:cosmology} that $\Mnu$ alters the height of the
peaks in the CMB power spectrum, and it will therefore be strongly
correlated with $\sigma_8$ which is an amplitude parameter. 

Effects on $n_s$ may be even more interesting. The significance of
$n_s<1$ is important, as $n_s\lesssim 1$ is a generic prediction of
most inflation models. In the case of $\mnue=0$eV, we find a
significance of $2.7\sigma$ for $n_s<1$, while this increases to
$3.8\sigma$ when applying a KATRIN prior in the case of
$\mnue=0.3$eV. This degeneracy between $\Mnu$ and $n_s$ stems from the
fact that increasing $\Mnu$ tends to increase the amplitude of the
acoustic peaks on scales smaller that the neutrino free streaming
scale. This can be compensated by decreasing $n_s$ which will give a
larger tilt on the primordial power spectrum.

Modifications in the distributions of $\Omega_m$ and
$\Omega_\Lambda$ are also evident. This happens since increasing $\Mnu$ will shift the
time of matter-radiation equality and thus amplify the acoustic
peaks, and this effect can be compensated by increasing $\Omega_m$
(and thus also reducing $\Omega_\Lambda$). 

\begin{figure}[htb]
\center
\includegraphics[width=14cm]{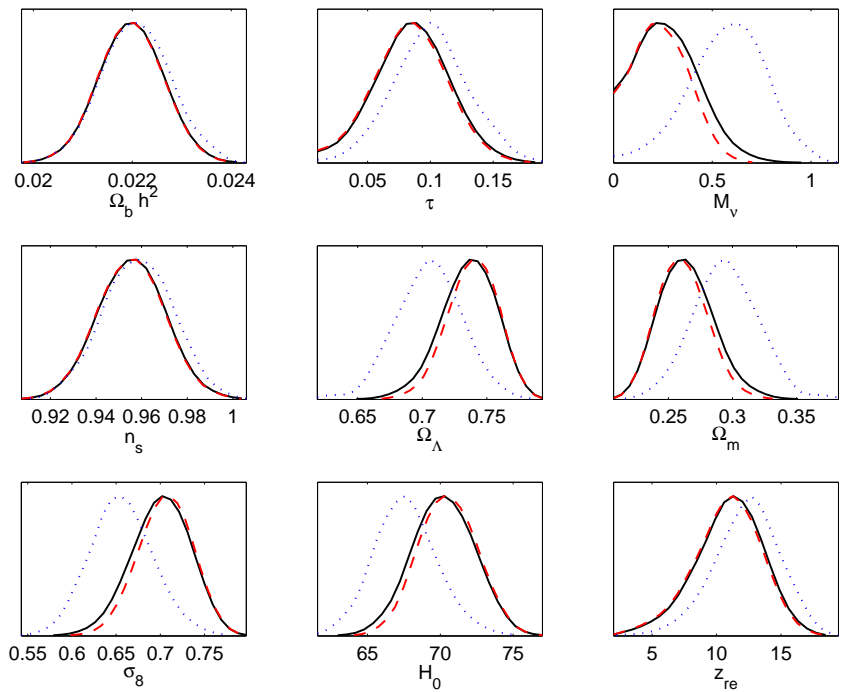}
\caption{\label{fig:2} Marginalized parameter distributions when using
  data from WMAP+LSS+SN1+HST+BBN+BAO+CMF (black solid lines), compared to the resulting distributions when
adding KATRIN data with $\mnue=0$eV (red dashed lines) and $\mnue=0.3$eV
(blue dotted lines).}
\end{figure}

Next, we added data from LSS, SN1, HST, BBN, BAO and CMF to the WMAP
data. Doing this, the limit on $\Mnu$ from cosmology alone is in the
same range as we get from KATRIN in the case of $\mnue=0$, thus we
would not expect a large effect from adding the KATRIN prior in this
case. If, however, we apply the $\mnue=0.3$eV scenario, we would
expect to see some effects. This is indeed the case, as we can see
from Figure \ref{fig:2} and Table \ref{tab:2}. 

\begin{table}[htb]
\center
\begin{tabular}{cccc}
\hline \hline
%Parameter & \multicolumn{3}{c}{WMAP} & \multicolumn{3}{c}{WMAP+LSS+SN1a+HST+BBN+CMF} \\
%\hline \hline
&  $\mnue = \textrm{free}$ & $\mnue=0$eV & $\mnue=0.3$eV \\
\hline
$\Omega_b h^2$ & $0.0220\pm0.0006$ & $0.0220\pm0.0006$ & $0.0221\pm0.0007$ \\
$h$ & $0.70\pm0.02$ & $0.71\pm+0.02$ & $0.68\pm0.02$\\
$\tau$ & $0.086\pm0.029$ & $0.084\pm0.028$ & $0.100\pm0.029$\\
$n_s$ & $0.956\pm0.014$ & $0.955\pm0.014$ & $0.958\pm0.015$\\
$\sigma_8$ & $0.70\pm0.03$ & $0.71\pm0.03$ & $0.66\pm0.04$ \\
$\Omega_\Lambda$ & $0.74\pm0.02$ & $0.74\pm0.02$ & $0.71\pm0.03$\\
$\Omega_m$ & $0.26\pm0.02$& $0.26\pm0.02$ & $0.29\pm0.03$ \\
$z_\textrm{re}$ & $10.9\pm2.6$ & $10.8\pm2.6$ & $12.3\pm2.5$\\
$\Mnu$ (eV) & $<0.55$ (95\%C.L.)& $<0.47$ (95\% C.L.)& $0.58^{+0.15}_{-0.17}$\\
\hline
\end{tabular}
\caption{\label{tab:2}Limits on cosmological parameters with different
  priors on the neutrino mass using the full range of data sets  WMAP+LSS+SN1a+HST+BBN+BAO+CMF. In the left column are the results from
having no priors on $\Mnu$ (black solid lines in Figure \ref{fig:1}),
the middle column shows the results when using the assumed KATRIN prior in
the case of $\mnue=0$eV (red dashed lines in Figure \ref{fig:1}), and
the rightmost column gives the results with an assumed KATRIN prior
for $\mnue=0.3$eV.}
\end{table}
The effect of adding a KATRIN prior with $\mnue=0.3$eV is most
pronounced for the $\Omega_m$, $\Omega_\Lambda$, $\sigma_8$ and
$h$. The shifts in the distributions can be explained by much of the
same arguments as in the case where we used WMAP data only. A
difference can be seen in the effect on $n_s$. When using only WMAP
data, a larger $\mnue$ pulled $n_s$ to lower values, while here $n_s$
is shifted to slightly larger values. This can be understood by the
effect of $\Mnu$ on the matter power spectrum, where a larger $\Mnu$
suppresses small scale fluctuations. This can be compensated by
increasing $n_s$. 

\subsection{Constraining $w$}

\begin{figure}[htb]
\center
\includegraphics[width=7cm]{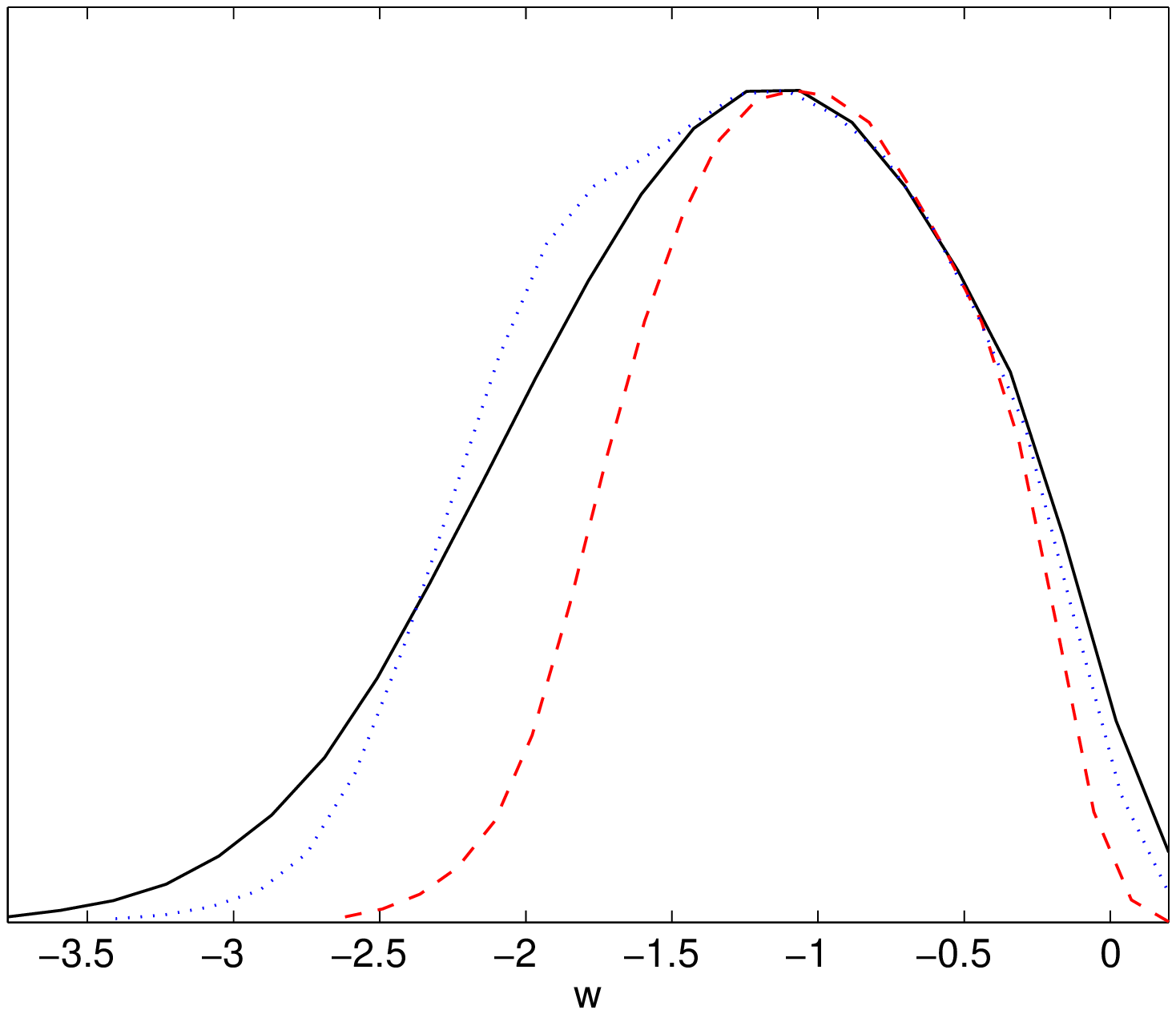}
\includegraphics[width=7cm]{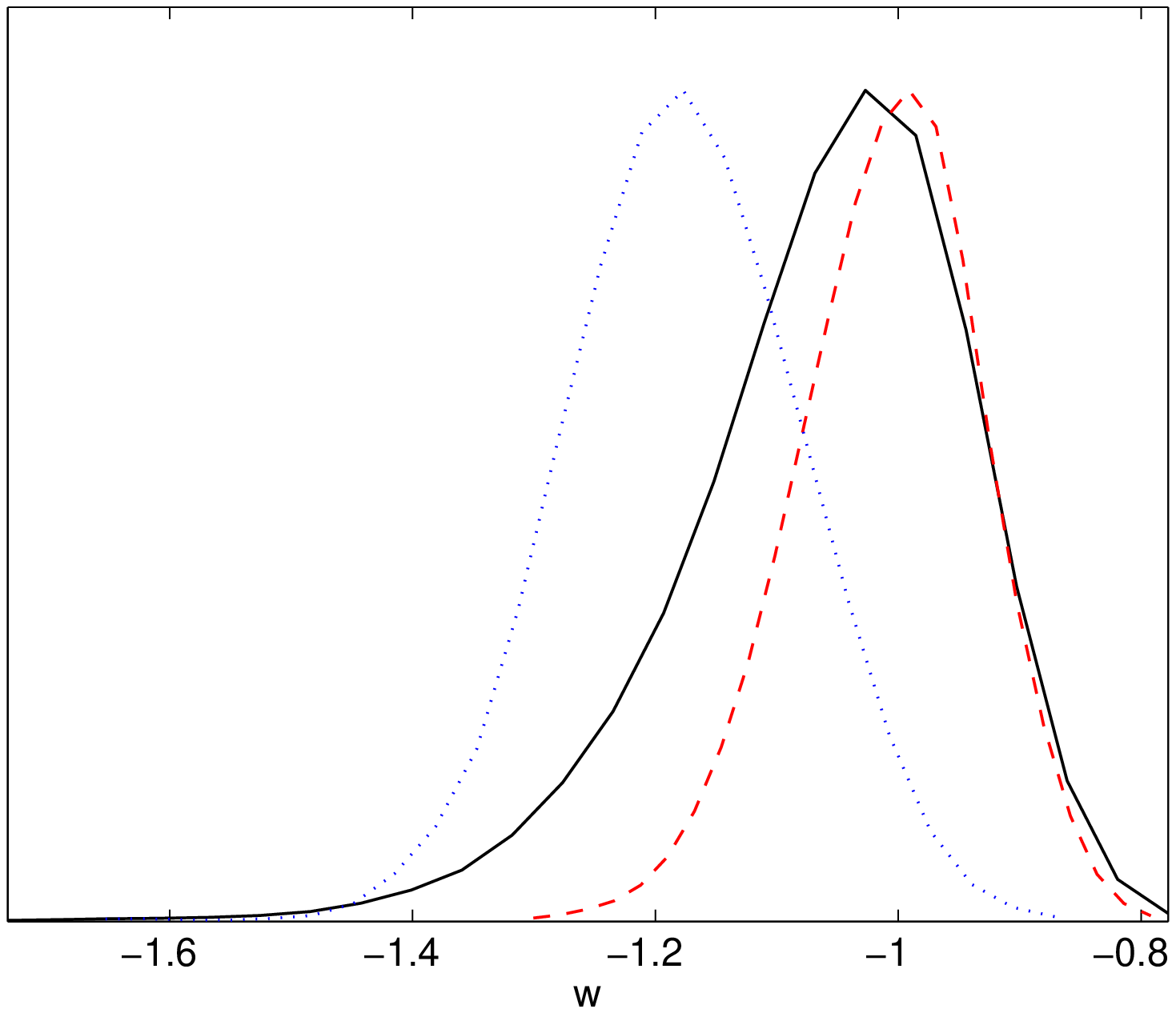}
\caption{\label{fig:3} Marginalized parameter distributions of w when using
  data from WMAP only (left panel) and WMAP+LSS+SN1a+HST+BBN+BAO+CMF(right
  panel). Black solid lines show the distributions for cosmological
  data only. This is compared to the resulting distributions when
adding KATRIN data with $\mnue=0$eV (red dashed lines) and $\mnue=0.3$eV
(blue dotted lines).}
\end{figure}

Next we redo the analysis, including $w$ as a free
parameter. The resulting constraints on $w$ can be seen in Figure
\ref{fig:3} and Table \ref{tab:3}.

\begin{table}[htb]
\center
\begin{tabular}{cccc}
\hline \hline
Cosmological data &  $\mnue = \textrm{free}$ & $\mnue=0$eV & $\mnue=0.3$eV \\
\hline
WMAP & $-1.3^{+0.5}_{-0.6}$ & $-1.1\pm0.4$ &  $-1.3\pm0.5$ \\
WMAP++ & $-1.06^{+0.10}_{-0.11}$ & $-1.01\pm0.07$ & $-1.18\pm0.09$\\
\hline
\end{tabular}
\caption{\label{tab:3}Limits on the equation of state of dark energy, $w$,  with different
  priors on the neutrino mass. The first row are the results from
  using WMAP data only. The second row comes from using the full range of data sets  WMAP+LSS+SN1a+HST+BBN+BAO+CMF.}
\end{table}

When using WMAP data only, the constraints on $w$ are relatively
weak. One usually needs to include e.g. SN1a or LSS data to get
tight constraints on this parameter, as the main effect of $w$ is to change the expansion history at
late times. We see, however, that the lower
limit on $w$ improves from $w>-2.5$ to $w>-1.8$ (95\% C.L.) when
including the KATRIN prior for $\mnue=0$eV. This happens since
increasing $w$ will decrease the late-time expansion rate of the
universe and thus shift the peaks in the CMB power spectrum to the
left. This can be compensated by reducing $\Mnu$, which will shift the
peaks back to the right.
 
When using the full range of cosmological data sets, the results
become more interesting, as the constraints on $w$ are tighter in this
case. Here, the most interesting effect occurs in the case of adding
a KATRIN prior for $\mnue=0.3$eV, which makes $w=-1$ excluded by
approximately 2$\sigma$. That means that a positive KATRIN detection
of $\mnue$ in this range, would be an indication of $w<-1$, which
would be very interesting from a cosmological point of view. 

The reason for this dependence on $\Mnu$, can partially be explained by
the shift of the peaks in the CMB power spectrum. But maybe more
important is the fact that a smaller $w$ will require a smaller
$\Omega_\Lambda$ to accommodate the accelerated expansion observed in
the SN1a data. This will in turn increase $\Omega_m$, which will allow
for larger neutrino masses, as explained earlier. 

\section{Goodness of fit}
\label{sec:chisq}

Another interesting issue is the change in $\chi^2$ when adding the
KATRIN priors. For each combination of cosmological model and data
sets, we have calculated the change in $\chi^2$ by
\bee
\Delta \chi^2 = 2 \ln \mathcal L(\textrm{no prior on } \Mnu) - 2 \ln \mathcal
L(\textrm{KATRIN prior on } \Mnu),
\ene
where $\mathcal L$ is the likelihood of the best-fit sample in each
combination of data sets on cosmological model. Thus $\Delta \chi^2$
is a measure of how much worse the fit to the data becomes by
introducing the KATRIN priors in the different cases. The resulting
values of $\Delta \chi^2$ are summarized in Table \ref{tab:chisq}.  

We see that in the case of $\mnue=0$eV, the changes in $\chi^2$ are
small compared to the models with no prior on $\Mnu$. This is not
surprising, as values of $\Mnu \approx 0$eV fit the cosmological data
very well. When using the full range of data sets in our 7 parameter
model, the situation becomes slightly worse, giving $\Delta \chi^2 =
-4.08$. Introducing $w$ as a free parameter in our model, the change
in $\chi^2$ by introducing the $\mnue=0.3$eV is not that severe
anymore, giving $\Delta \chi^2=-1.59$. This can be understood by the
well-known degeneracy between $\Mnu$ and $w$. 

\begin{table}[htb]
\center
\begin{tabular}{llccc}
\hline \hline
Cosmological data &  Model & $\Delta \chi^2 (\mnue=0$eV) & $\Delta
\chi^2 (\mnue=0.3$eV) & $N_\textrm{par}$\\
\hline
WMAP & $\Lambda$CDM + $\Mnu$ & -0.01 & -0.18 & 7     \\
WMAP++ & $\Lambda$CDM + $\Mnu$ & -0.02 & -4.08 & 7    \\
WMAP & $\Lambda$CDM + $\Mnu + w$ & -0.13 & -0.01 & 8      \\
WMAP++ & $\Lambda$CDM + $\Mnu + w$ & -0.07 & -1.59 & 8    \\
\hline
\end{tabular}
\caption{\label{tab:chisq} $\Delta \chi^2$ of the models with a KATRIN
  prior
  on $\Mnu$ relative to the models with no prior on $\Mnu$. WMAP++
  refers to the analysis with using all data sets
  WMAP+LSS+SN1a+HST+BBN+BAO+CMF as described in the text. $N_\textrm{par}$
  refers to the number of free parameters in the models.}
\end{table}

\section{Discussion and conclusions}
\label{sec:conclusions}
In this paper we have investigated whether constraints on $\mnue$ from
the KATRIN experiment will affect our knowledge on cosmological
parameters. This has been done for two scenarios, one where
$\mnue=0$eV, and one where $\mnue=0.3$eV. We have carried out the analysis
both with a simple 7-parameter model with a cosmological constant, and
extending the parameter space to include the equation of state for
dark energy, $w$, as a free parameter.

When using WMAP data only, we find that knowledge from the KATRIN
experiment will contribute significantly to constrain a wide range of cosmological
parameters, regardless of which of the $\mnue$ scenarios we use. For
instance will the significance of $n_s<1$ depend on what KATRIN tells
us about $\mnue$. Other parameters that are sensitive to the value of
$\mnue$ are $\Omega_m$, $\Omega_\Lambda$, $\sigma_8$ and $H_0$. 

Adding more cosmological data sets, both from SN1a, galaxy catalogues
and other priors, the situation changes a bit. In this case $\Mnu$
is strongly constrained from above by cosmology alone, such that an
additional KATRIN prior in the case of $\mnue=0$eV has little effect on
our cosmological parameter constraints. However, if KATRIN measures a
neutrino mass of $\mnue=0.3$eV, there will be significant shifts in
several of the parameter distributions. One should also note that
several of the extra cosmological data sets added here may be affected
by uncontrolled systematics (see \cite{kristiansen:2007}). Therefore,
having cosmological constraints from WMAP+KATRIN without any
additional cosmological data sets will be interesting regardless of
the possibility to add other cosmological data sets to obtain similar
results.

In the case of $w$ the most interesting result occurs in the scenario
of a KATRIN defection of $\mnue=0.3$eV and using the full range of
data sets. In this case, $w<-1$ is favored at a $2\sigma$ level. It should
also be mentioned that there are degeneracies between parameters from
different cosmological inflation models and neutrino masses (see
\cite{hamann:2006}). This means that a KATRIN prior on $\Mnu$ will be
important also for constraining inflationary models. 

To conclude, we find that the expected limits on $\mnue$ from KATRIN,
will be a useful input to constrain cosmological models, regardless of
the value of $\mnue$. If KATRIN detects a non-zero value of $\mnue$,
this would be especially interesting.  

%%%%%%%%%%%%%%%%%%%%%%%%%%%%%%%%%%%%%%%%%%%%%%%%%%%%%%%%%%%%%%%%%%%%%%
\section*{Acknowledgments} %%%%%%%%%%%%%%%%%%%%%%%%%%%%%%%%%%%%%%%%%%%
%%%%%%%%%%%%%%%%%%%%%%%%%%%%%%%%%%%%%%%%%%%%%%%%%%%%%%%%%%%%%%%%%%%%%%
We wish to thank Ole H{\o}st, Klaus Eitel and Hans Kristian Eriksen 
for useful comments and suggestions. 
The work of {\O}E is supported by the Research Council of Norway, project 
number 162830.

%%%%%%%%%%%%%%%%%%%%%%%%%%%%%%%%%

%\bibstyle{unsrt.bst}
%\bibliography{cites}

\end{document}